# Realization of a two-dimensional checkerboard lattice in monolayer Cu$_2$N

Xuegao Hu,[1,3,#] Run-Wu Zhang,[2,#] Da-Shuai Ma,[4,#] Zhihao Cai,[1,3] Daiyu Geng,[1,3] Zhenyu Sun,[1,3] Qiaoxiao Zhao,[1,3] Jisong Gao,[1,3] Peng Cheng,[1,3] Lan Chen,[1,3,5] Kehui Wu,[1,3,5,6]* Yugui Yao,[2]* Baojie Feng,[1,3,6]*

[1]Institute of Physics, Chinese Academy of Sciences, Beijing, 100190, China

[2]Centre for Quantum Physics, Key Laboratory of Advanced Optoelectronic Quantum Architecture and Measurement (MOE), School of Physics and Beijing Key Lab of Nanophotonics Ultrafine Optoelectronic Systems, School of Physics, Beijing Institute of Technology, Beijing 100081, China

[3]School of Physical Sciences, University of Chinese Academy of Sciences, Beijing, 100049, China

[4]Institute for Structure and Function & Department of Physics, Chongqing University, Chongqing 400044, China

[5]Songshan Lake Materials Laboratory, Dongguan, Guangdong, 523808, China

[6]Interdisciplinary Institute of Light-Element Quantum Materials and Research Center for Light-Element Advanced Materials, Peking University, Beijing, 100871, China

[#]These authors contributed equally to this work.

[*]Corresponding author. E-mail: khwu@iphy.ac.cn; ygyao@bit.edu.cn; bjfeng@iphy.ac.cn.

## Abstract

**Two-dimensional checkerboard lattice, the simplest line-graph lattice, has been intensively studied as a toy model, while material design and synthesis remain elusive. Here, we report theoretical prediction and experimental realization of the checkerboard lattice in monolayer Cu$_2$N. Experimentally, monolayer Cu$_2$N can be realized in the well-known N/Cu(100) and N/Cu(111) systems that were previously mistakenly believed to be insulators. Combined angle-resolved photoemission spectroscopy measurements, first-principles calculations, and tight-binding**

**analysis show that both systems host checkerboard-derived hole pockets near the Fermi level. In addition, monolayer $Cu_2N$ has outstanding stability in air and organic solvents, which is crucial for further device applications.**

Two-dimensional (2D) materials with nontrivial lattice structures can host rich physical properties, which are protected by the symmetries of the lattices. For example, the honeycomb lattice, such as graphene[1] and silicene[2], is manifested by linearly dispersing Dirac cones in their electronic band structures. On the other hand, in 2D line-graph lattices, including kagome and checkerboard lattices[3,4], symmetry constraints lead to destructive interference of Bloch waves, giving rise to compact localization of wave functions in the real space or topological flat bands in the momentum space. Distinct from atomic and molecular orbitals that are also flat in the momentum space, topological flat bands contain extended states, and electrons on these states have quenched kinetic energy and diverging effective masses. Therefore, 2D line-graph lattices provide a fertile playground to study strongly correlated physics, including fractional quantum Hall effect[5], unconventional superconductivity[6], and Wigner crystallization[7].

The rich physical properties in line-graph lattices have spurred intense research interest in realizing candidate materials for practical applications. For example, three-dimensional crystals constructed from kagome lattices, exemplified by $Co_3Sn_2S_2$ and $A$V$_3$Sb$_5$ ($A$ = K, Rb, Cs), have been intensively studied in the recent few years. Many intriguing properties have been discovered, including the giant anomalous Hall effect, chiral charge density wave, and unconventional superconductivity[8–15]. However, the intrinsic kagome bands in most of these materials are seriously deformed because of the complex bonding configurations and nonnegligible interlayer coupling in real materials. From this viewpoint, it is highly desirable to realize line-graph lattices with simpler atomic structures in monoatomic layer materials. As the simplest line-graph lattice, the checkerboard lattice has only two atoms in each unit cell and has received significant attention from theorists[16–21]. However, both material prediction and realization of the checkerboard lattice are challenging because of the simultaneous requirements of a fourfold symmetric structure and twofold symmetric hoppings at each lattice site.

In this work, we demonstrate that monolayer $Cu_2N$ has an ideal checkerboard lattice and can be grown on Cu(100) and Cu(111) by molecular beam epitaxy (MBE). The growth of nitrogen on Cu was reported decades ago[22–26], but the surface nitride layers were interpreted as surface adsorption systems in previous works. Early scanning tunneling spectroscopy (STS) measurements claimed that the surface nitride layer insulates with a band gap exceeding 4 eV. As a result, nitrogen adsorption on Cu was frequently used to decouple the hybridization of single atoms or molecules with the metallic Cu substrates[27,28]. To date, detailed research on the intrinsic properties of monolayer $Cu_2N$, especially its topological nature derived from the unique lattice symmetry, is still lacking.

Here, we systematically study the electronic structure and topological properties of monolayer $Cu_2N$ by combined first-principles calculations, tight-binding (TB) model analysis, and angle-resolved photoemission spectroscopy (ARPES) measurements. Our theoretical calculations show that freestanding $Cu_2N$ is thermodynamically stable with all Cu and N atoms coplanar. The stability of freestanding $Cu_2N$ makes it a 2D material compared to the kagome layer in most kagome materials. The unique checkerboard lattice gives rise to a topological flat band and a saddle point in the proximity of the Fermi level and a hole-like band at each M point. Experimentally, we synthesized monolayer $Cu_2N$ on Cu(100) and Cu(111) by MBE and observed the checkerboard-derived hole-like bands by ARPES measurements. The topological flat band and saddle point of monolayer $Cu_2N$ are hybridized with the substrates, as confirmed by our ARPES measurements and first-principles calculations. In addition, we proved the outstanding stability of monolayer $Cu_2N$ in air and various organic solvents, which is crucial for further device applications.

First, we briefly review the graph theory that can help to understand 2D frustrated lattices. A line graph is constructed by connecting the centers of edges sharing a common vertex in the original graph. The electronic Hamiltonian of a line-graph lattice is equivalent to the Laplacian operator, which has a constant eigenvalue, giving rise to flat bands[29]. For instance, the kagome lattice is the line graph of the honeycomb lattice, and the checkerboard lattice is the line graph of the square lattice[21]. Schematic drawings of the square and checkerboard lattices are shown in Fig. 1(a) and 1(b). The checkerboard lattice has two different hoppings: $t_1$ between A and B sites and $t_2$ between A and C sites. The hopping between A and D sites is forbidden, which results in twofold symmetric hoppings at the A site. The Hamiltonian eigenfunctions have opposite amplitudes at A and B sites. As a result, electronic states are geometrically confined within a single square, as indicated by the grey squares in Fig. 1(b). The real-space electronic localization leads to a flat band in the reciprocal space.

The unique symmetry requirement of the checkerboard lattice is reminiscent of the $c(2\times2)$ superstructure in surface reconstructions. Consider a square lattice with only the nearest neighbor hopping $t_1$ and the next-nearest neighbor hopping $t_2$ has vanishing amplitudes, as shown by the orange balls in Fig. 1(b). If we add an atom on each intersection of $t_2$, this atom will bridge the next-nearest neighbor sites and leads to a finite value of $t_2$, forming a checkerboard lattice. According to this idea, we constructed monolayer $Cu_2N$, as shown in Fig. 1(c). The crystal structure of $Cu_2N$ belongs to the space group *P4/mmm* (No.123), and the 1a and 2f Wyckoff positions are occupied by the N and Cu atoms, respectively. After structural optimization, all atoms in freestanding $Cu_2N$ are coplanar, as shown in the bottom panel of Fig. 1(c). Our extensive ground state search indicated that monolayer $Cu_2N$ with the space group of *P4/mmm* is the ground state of stoichiometric $Cu_2N$, as shown in Supplementary Fig. S1. The calculated phonon spectrum of freestanding $Cu_2N$ is shown in Fig. 1(d), and no imaginary frequencies were observed, which confirmed the dynamic stability of freestanding $Cu_2N$. The thermostability of freestanding $Cu_2N$ was further confirmed by our *ab initio* molecular dynamics simulation, as shown in Supplementary Fig. S2.

Figure 1(e) shows the DFT-calculated band structures of monolayer $Cu_2N$. At the

M point, we observed a hole-like band capped with a relatively flat band in the proximity of the Fermi level, analogous to the characteristic band structure of the checkerboard lattice. However, the flat band becomes slightly dispersive away from the M point, which indicates a slight deviation from the ideal model. Our orbital-projected band structure calculations (Supplementary Fig. S3) show that the flat and parabolic bands are mainly contributed by the Cu $d_{xy}$ orbitals, which proves that the checkerboard lattice is constructed by the Cu atoms. On the other hand, the N atoms contribute a saddle point at the X point, as indicated by the red arrow in Fig. 1(e).

To understand the checkerboard-derived flat and parabolic bands, we build a TB model by putting the Cu $d_{xy}$ orbitals at A and B sites [Fig. 1(b)]. The Bloch basis can be described as:

$$|\phi_\alpha(k)\rangle = \sum_R e^{iR\cdot k}|\alpha, R\rangle_{\alpha=A,B}, \qquad (1)$$

the TB Hamiltonian can be expressed as:

$$H(k) = \begin{pmatrix} e^{-ik_x}(1+e^{2ik_x})t_2 & e^{-ik_x}(1+e^{ik_x})(1+e^{ik_y})t_1 \\ e^{ik_x}(1+e^{-ik_x})(1+e^{-ik_y})t_1 & e^{-ik_y}(1+e^{2ik_y})t_2 \end{pmatrix}, \qquad (2)$$

where $t_1$ and $t_2$ are the hopping integrals shown in Fig. 1(b). When $t_1 = t_2$, the Hamiltonian $H(k)$ describes an ideal checkerboard lattice that hosts a parabolic band at the M point and a flat band in the whole Brillouin zone (BZ), as shown in Fig. 1(f)[16,17]. In real materials, $t_1$ is not exactly equal to $t_2$, which will lead to the geometrical obstruction of the checkerboard lattice. As a result, the otherwise completely flat band becomes dispersive away from the M point. Figure 1(g) shows the energy spectrum when $t_1 = 0.8t_2$, which qualitatively agrees with our DFT calculation results. Notably, the off-diagonal elements of Eq. 2 vanish along the BZ boundary. Thus, Eq. 2 is deduced to the model of a checkerboard lattice, and one of the bands is completely flat, with the corresponding energy being -$2t_2$.

The calculated band structures of $Cu_2N$ with superposition of contributions from Cu $d_{xy}$ orbitals are shown in Fig. 1(h) (without SOC). By comparing the band structure of monolayer $Cu_2N$ with the energy spectrum of the checkerboard lattice, we find that monolayer $Cu_2N$ is a perturbed checkerboard lattice. Because of the perturbation that originates from the difference between $t_1$ and $t_2$, the flat band becomes slightly dispersive but keeps flat along the boundary of the BZ. When SOC is considered, the band touching at the M point will be gapped out. To determine the topological properties of $Cu_2N$, we calculated the $Z_2$ invariant based on the method proposed by Fu and Kane[30]. The calculation result is $Z_2 = 1$, which indicates the nontrivial topology of the flat band.

To experimentally verify the topological electronic structure of monolayer $Cu_2N$, we synthesized the samples on Cu(100) and Cu(111) by MBE and performed ARPES measurements. The sharp low-energy electron diffraction (LEED) patterns shown in Fig. 2 indicate the high quality of our samples. For $Cu_2N$/Cu(100), the N atoms sit at the $c(2\times2)$ hollow sites of the topmost Cu layer[27], forming a commensurate $Cu_2N$ monolayer on Cu(100), as shown in Figs. 2(d) and 2(h). Our first-principles calculations show that monolayer $Cu_2N$ remains flat on Cu(100) after structural relaxation. On the other hand, monolayer $Cu_2N$ can also grow on Cu(111). However, the threefold symmetry of Cu(111) leads to the coexistence of three equivalent $Cu_2N$ domains rotated by 120º. The successful growth of monolayer $Cu_2N$ on two different Cu facets, including an incommensurate Cu(111) substrate, indicates that monolayer $Cu_2N$ is a stable 2D material, in agreement with our phonon spectrum calculation results.

To test the chemical stabilities of monolayer $Cu_2N$, we transferred the as-prepared samples from vacuum to air. After 5 minutes, the samples were reloaded into the MBE chamber, followed by mild degassing for 30 minutes. Surprisingly, the LEED patterns of $Cu_2N$ survive, and their sharpness is comparable to the original ones, as shown in Figs. 2(c) and 2(g). In addition, we found that $Cu_2N$ is also stable in various solvents, including alcohol, acetone, and deionized water, as shown in Supplementary Fig. S4. The stability of $Cu_2N$ has also been confirmed by our scanning tunneling microscope measurements, as shown in Supplementary Fig. S5. These results show that monolayer $Cu_2N$ has outstanding chemical stability, which is crucial for its device applications.

The chemical stability of Cu$_2$N might originate from the positive valence states of Cu atoms, which makes it resistant to further oxidization in ambient conditions.

We then performed ARPES measurements to study the electronic structure of monolayer Cu$_2$N. Figures 3(a)-3(d) show the constant energy contours (CECs) of Cu$_2$N/Cu(100) at different binding energies. On the Fermi surface [Fig. 3(a)], there is a dot-like feature at each M point of Cu$_2$N, which is absent in pristine Cu(100) [Supplementary Fig. S6]. With increasing binding energies, each dot-like feature evolves into a closed pocket and becomes larger at higher binding energies, indicating a hole-like band at each M point of Cu$_2$N. Similar results have also been observed in Cu$_2$N/Cu(111), as shown in Figs. 3(e)-3(h). However, the coexistence of three equivalent domains gives rise to 12 dot-like features on the Fermi surface.

Notably, pristine Cu(100) and Cu(111) do not have similar hole-like bands given any momentum or energy shift, which rules out the band-folding effects or rigid shifts of the bulk bands of Cu. In addition, each pocket at high binding energies is fourfold symmetric on both Cu(100) and Cu(111), which agrees with the fourfold symmetry of Cu$_2$N. Therefore, we conclude that the hole-like band at the M point originates from the surface Cu$_2$N layer instead of the substrate.

ARPES intensity plots along Cuts 1-4 [marked in Figs. 3(a) and 3(e)] are displayed in Fig. 4. A hole-like band at the *M* point is observed. The band maximum is slightly above the Fermi level, as indicated by the red dashed lines in Figs. 4(a) and 4(c). The observation of hole-like bands agrees with the evolution of the CECs in Fig. 3. The Fermi velocity along Cut 1 is slightly higher than that along Cut 3, which is caused by the fourfold symmetry of each pocket. We did not observe other bands within 1.5 eV below the Fermi level. The hole-like band at each M point is reminiscent of the calculation results of freestanding Cu$_2$N [Figs. 1(g) and 1(h)], but the flat band above was not observed. Our calculated band structures including the substrate agree well with our experimental results, as shown in Figs. 4(b), 4(d), and Supplementary Fig. S7. Therefore, we conclude that the disappearance of the flat band is due to the hybridization of Cu$_2$N with the Cu substrate.

Finally, we discuss the discrepancy between our results with previous STS results. Previous STS measurements reveal a large band gap exceeding 4 eV[25], while our combined ARPES and theoretical studies show that monolayer $Cu_2N$ is a metal. As is well known, the STS signal acquired with the constant height mode in previous works is only sensitive to electronic states at the $\Gamma$ point. Since the hole-like bands in monolayer $Cu_2N$ are located at the M points, they have little contribution to the STS signal, which can explain the gap feature in previous STS data[31,32]. On the other hand, because of the excellent chemical stability, $Cu_2N$ is unlikely to hybridize with most adsorbing atoms or molecules and can thus be used to avoid coupling of single atoms or molecules with the substrates[27,28].

To conclude, our combined theoretical calculations and ARPES measurements confirmed the realization of the checkerboard lattice in monolayer $Cu_2N$. The flat band and saddle point of freestanding $Cu_2N$ are located near the Fermi level, which might give rise to intriguing properties. We prepared the samples on Cu(100) and Cu(111) and observed checkerboard-derived bands in both systems, although some bands are hybridized with the substrates because of the substrate-overlayer interaction. Our work may stimulate further theoretical and experimental efforts to explore the exotic properties of the checkerboard lattice.

**Methods**

The sample preparation and measurements were performed in an ultrahigh vacuum system with a base pressure of ~$1\times10^{-8}$ Pa. Cu(100) and Cu(111) surfaces were cleaned by repeated Ar ion sputtering and annealing cycles. Monolayer $Cu_2N$ was grown on Cu(100) and Cu(111) by nitrogen ion bombardment at room temperature for 10 min, followed by annealing to 600 K. The cleanliness of the substrates and the structure of $Cu_2N$ were confirmed by LEED and ARPES measurements. ARPES measurements were performed with a SPECS PHOIBUS 150 electron energy analyzer and a He discharge lamp (He I$\alpha$ light) at 30 K.

First-principles calculations were applied based on the Vienna ab-initio simulation code[33] within the generalized gradient approximation (GGA) of the Perdew, Burke, and Ernzerhof exchange-correlation potential[34]. Thereinto, the energy cutoff for the plane-wave basis set was set to be 550 eV, and a $k$-mesh of 20×20×1 was adopted in self-consistent-field calculations. A vacuum region of about 20 Å was added along the z direction to avoid the interaction between adjacent parts. All the atomic positions and the lattice constants were fully relaxed until the maximal forces were less than 0.001 eV/Å, and the convergence criteria for energy was set to be $10^{-6}$ eV. To optimize the structure of different phases of $Cu_2N$ on Cu(111), we employed the SCAN+rVV10 approach[35]. The phonon spectra were calculated using the finite displacement method implemented in PHONOPY[36]. The ground state search for $Cu_2N$ was performed using the particle swarm optimization implemented in the CALYPSO code[37], which efficiently identifies the ground state and metastable structures based on the given chemical compositions.

**Associated Content**

**Supporting Information**

Calculated band structures of monolayer $Cu_2N$; LEED patterns of $Cu_2N$/Cu(100) after exposure to alcohol acetone mixture and deionized water; ARPES measurements of pristine Cu(100) and Cu(111); Calculated band structures of $Cu_2N$/Cu(100) with different layers of substrate.

**Acknowledgments**

This work was supported by the Ministry of Science and Technology of China (Grants No. 2018YFE0202700, No. 2020YFA0308800, and No. 2021YFA1400502), the National Natural Science Foundation of China (Grants No. 11974391, No. 11825405, No. 1192780039, No. 12234003, No. 12061131002, No. 12204074, and No. U2032204), the International Partnership Program of Chinese Academy of Sciences (Grant No. 112111KYSB20200012), the Strategic Priority Research Program of

Chinese Academy of Sciences (Grants No. XDB33030100 and No. XDB30000000), the CAS Project for Young Scientists in Basic Research (Grant No. YSBR-047), the China National Postdoctoral Program for Innovative Talent (Grant No. BX20220367), and the Beijing Institute of Technology Research Fund Program for Young Scholars.

**Author Contributions.**

B.F. conceived the research. X.H., Z.C., D.G., Z.S., Q.Z., J.G., and B.F. prepared the samples and performed ARPES experiments; R.Z., D.M., and Y.Y performed theoretical calculations and analysis; all authors contributed to the discussion of the data and writing of the manuscript.

**Notes**

The authors declare no competing interests.

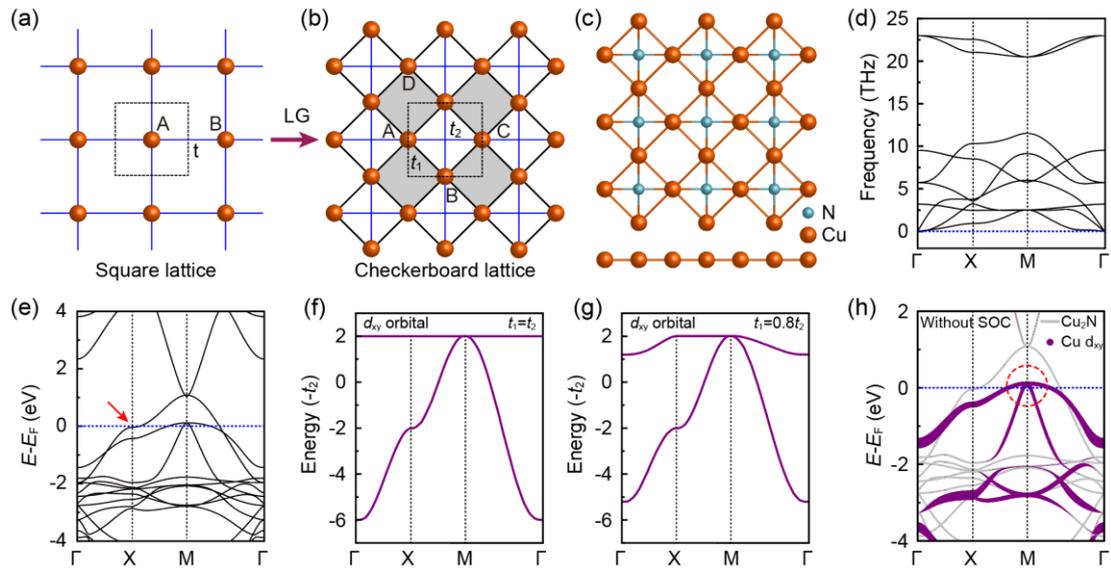

**Figure 1. Electronic structures of the checkerboard lattice and monolayer $Cu_2N$.** (a,b) Schematic drawing of the square and checkerboard lattice. One can obtain the checkerboard lattice by applying the line graph operation on the square lattice. Orange balls represent lattice sites; dashed lines indicate the unit cells; solid lines indicate allowed intersite hoppings. Grey squares indicate real-space electronic localization in the checkerboard lattice. (c) Schematic drawing of the atomic structure of monolayer $Cu_2N$. Orange and cyan balls represent Cu and N atoms, respectively. (d) Calculated phonon spectrum of freestanding $Cu_2N$. (e) Calculated band structures of freestanding $Cu_2N$ without SOC. The red arrow indicates a saddle point derived from the N atoms. (f,g) Band structures of the checkerboard lattice calculated using the $d_{xy}$ orbital with different hopping strengths: $t_1=t_2$ and $t_1=0.8t_2$, respectively. (h) Calculated band structures of freestanding $Cu_2N$ and with superposition of contributions from Cu $d_{xy}$ orbitals. The red dashed circle highlights the band touching that can be gapped out by SOC.

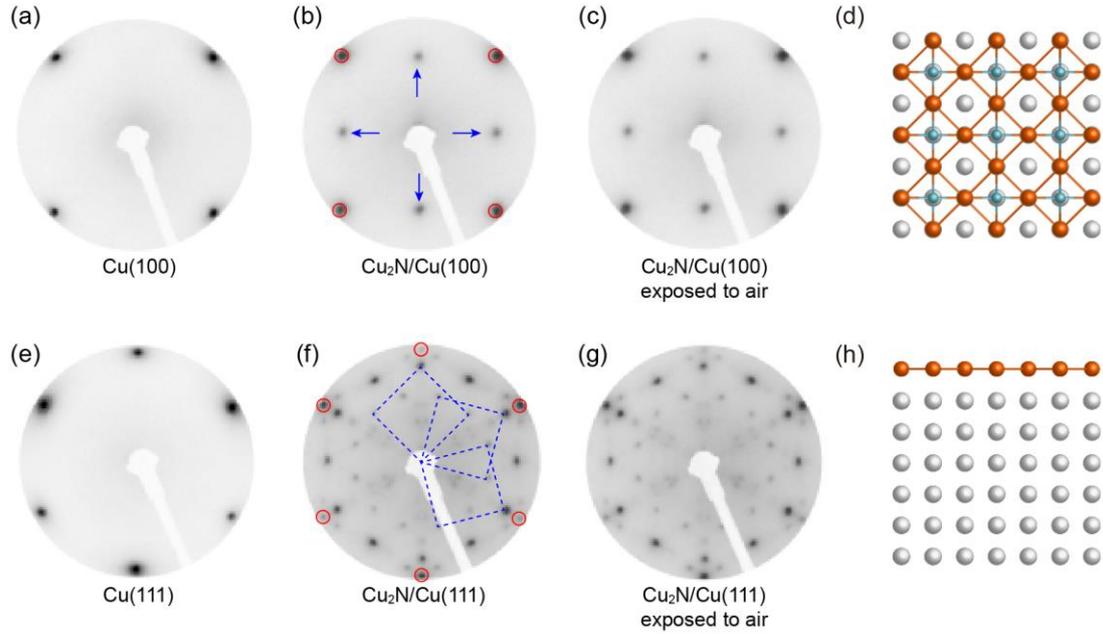

**Figure 2. Growth and stability of monolayer Cu$_2$N on Cu(100) and Cu(111).** (a,b) LEED patterns of Cu(100) and Cu$_2$N/Cu(100), respectively. Red circles indicate the diffraction spots of Cu(100). Blue arrows indicate the diffraction spots of monolayer Cu$_2$N, which forms a $(\sqrt{2} \times \sqrt{2})R45°$ superstructure with respect to Cu(100). (c) LEED patterns of Cu$_2$N/Cu(100) after exposure to air and degassing in vacuum. (e-g) The same as (a-c) but for Cu(111) substrates. Blue dashed squares in (f) indicate three equivalent Cu$_2$N domains. Incident electron energies are 45 eV for (a-c) and 55 eV for (e-g). (d,h) Top and side views of the atomic structure of Cu$_2$N/Cu(100), respectively.

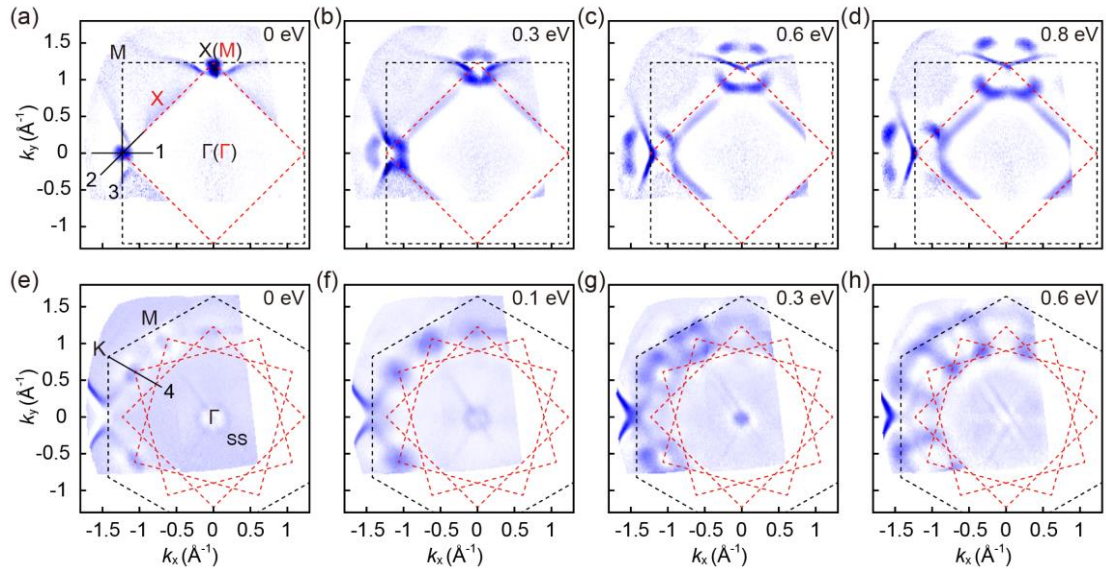

**Figure 3. ARPES intensity maps of the constant energy contours (CEC) of monolayer Cu$_2$N.** (a)-(d) Second-derivative CECs of Cu$_2$N/Cu(100) at binding energies of 0, 0.3, 0.6, and 0.8 eV, respectively. The X point of Cu(100) coincides with the M point of Cu$_2$N. (e)-(h) Second-derivative

CECs of $Cu_2N$/Cu(111) at binding energies of 0, 0.2, 0.4, and 0.6 eV, respectively. Black and red dashed lines indicate the BZs of Cu(100) and $Cu_2N$, respectively. There are three equivalent domains of $Cu_2N$ on Cu(111).

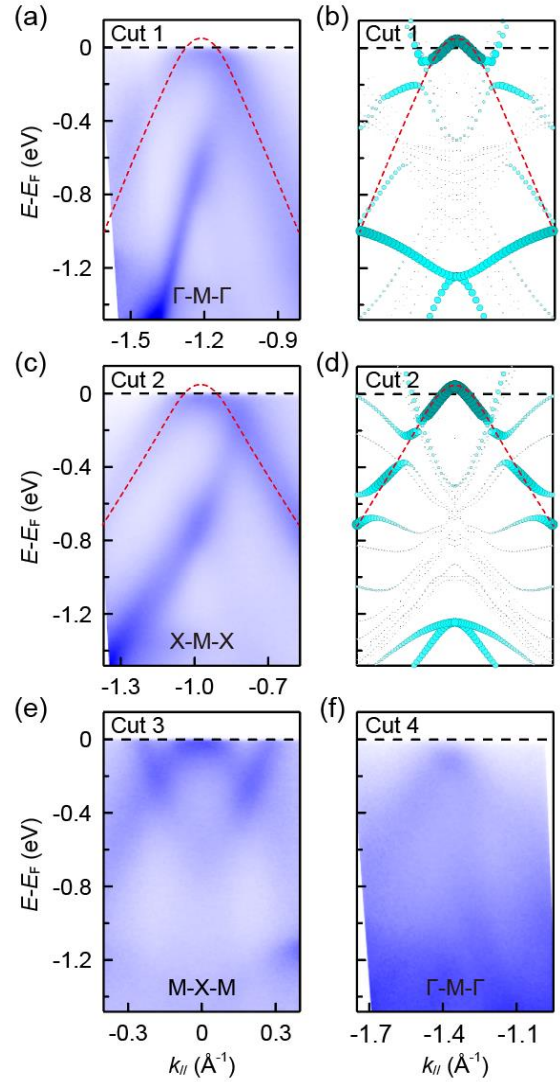

**Figure 4. ARPES intensity plots along high-symmetry momentum cuts.** (a,c,e,f) ARPES intensity plots along Cuts 1-4 indicated in Figs. 3(a) and 3(e). (b,d) Calculated band structures of $Cu_2N$/Cu(100) projected to the $Cu_2N$ layer along Cuts 1 and 2, respectively. The chemical potential was shifted 0.22 eV toward higher binding energy for better comparison with experimental results. Red dashed lines in (a-d) are guides to the eye.

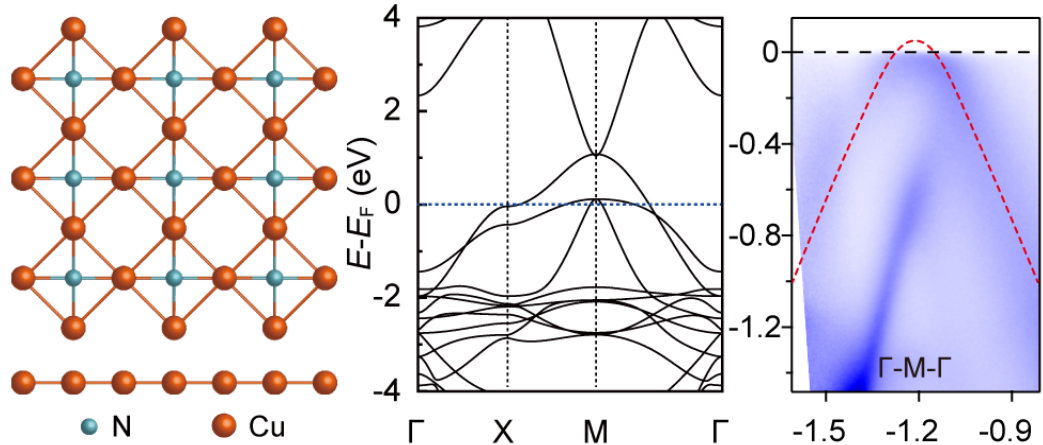

TOC Graphic